\documentclass[journal=jacsat,manuscript=article]{achemso}

\usepackage[version=3]{mhchem} 
\usepackage{algorithm}
\usepackage{amsmath}
 \usepackage{graphicx}
 \usepackage{subcaption}
 \usepackage{pdfpages}
\usepackage{algpseudocode} 
\usepackage{siunitx}
\usepackage{braket}
\usepackage{multirow}
\usepackage{mathpazo}
\usepackage{mathrsfs}  
\usepackage{mathtools}
\usepackage{commath}
\usepackage[section]{placeins}

\newcommand{\rvline}{\hspace*{-\arraycolsep}\vline\hspace*{-\arraycolsep}}
\DeclareMathOperator\erf{erf}
\DeclareMathOperator\trace{Tr}

\author{Pier Paolo Poier}
\affiliation{Sorbonne Universit\'e, LCT, UMR 7616 CNRS, Paris, France}
\email{pier.poier@sorbonne-universite.fr}
\author{Olivier Adjoua}
\affiliation{Sorbonne Universit\'e, LCT, UMR 7616 CNRS, Paris, France}
\author{Louis Lagardère}
\affiliation{Sorbonne Universit\'e, LCT, UMR 7616 CNRS, Paris, France}
\alsoaffiliation{Sorbonne Universit\'e, IP2CT, FR 2622 CNRS, Paris, France}
\author{Jean-Philip Piquemal}
\affiliation{Sorbonne Universit\'e, LCT, UMR 7616 CNRS, Paris, France}
\alsoaffiliation{The University of Texas at Austin, Department of Biomedical Engineering, TX, USA}
\email{jean-philip.piquemal@sorbonne-universite.fr}

\title[An \textsf{achemso} demo]
  {Generalized Many-Body Dispersion Correction through Random-phase Approximation for Chemically Accurate Density Functional Theory 
  }

\abbreviations{IR,NMR,UV}

\begin{document}


\begin{abstract}
We extend our recently proposed Deep Learning-aided many-body dispersion (DNN-MBD) model to quadrupole polarizability (Q) terms using a generalized Random Phase Approximation (RPA) formalism, thus enabling the inclusion of van der Waals contributions beyond dipole. The resulting DNN-MBDQ model only relies on \emph{ab initio}-derived quantities as the introduced quadrupole polarizabilities are recursively retrieved from dipole ones, in turn modelled via the Tkatchenko-Scheffler method. A transferable and efficient deep-neuronal network (DNN) provides atom in molecule volumes, while a single range-separation parameter is used to couple the model to Density Functional Theory (DFT). Since it can be computed at a negligible cost, the DNN-MBDQ approach can be coupled with DFT functionals such as PBE,PBE0 and B86bPBE (dispersionless). The DNN-MBQ-corrected functionals reach chemical accuracy while exhibiting lower errors compared to their dipole-only counterparts.







\end{abstract}

\section{Introduction}

The importance of modelling matter through computer simulations has risen tremendously in the past decades in virtue of both the theoretical achievements and the advent of mass-produced computers whose performances have increased exponentially. Despite these tremendous achievements, the exact solution of the non-relativistic electronic Schr\"odinger equation remains out of reach for multi-electron systems and different approximations have been introduced in order to model and tackle systems of chemical relevance. In particular, Kohn-Sham Density Functional Theory (KS-DFT)
has established itself as the most widely used electronic structure method being the cheapest way for introducing electronic correlation. KS-DFT is based on the idea of evaluating the kinetic energy from a single Slater determinant, thus assuming the electrons being non-interacting. The difference between the true and KS kinetic energies as well as the difference between the true electronic and exchange interaction ones is essentially embedded and modelled by the exchange correlation functional, key quantity in KS-DFT, however, in practice unknown.\\
The different approaches taken in modelling the exchange-correlation functional define the plethora of KS-DFT variants which, however, are mainly capable of capturing local correlation effects. Dispersion interactions, on the other hand, are rooted in the long-range electron correlation that can clearly not be captured by the intrinsic locality of common exchange correlation functionals. This inadequacy of KS-DFT in modelling non-covalent dispersion interactions, ubiquitous in nature and materials, has risen the attention towards the development of dispersion correction models\cite{grimmechemrev}, the most widely used being based on empirical pairwise terms, Eq.\eqref{london}.\cite{grimme} 
\begin{equation}
    \label{london}
    \mathcal{E}_\text{disp}=\sum_{i>j}^N-\frac{C_6^{ij}}{R^6_{ij}}
\end{equation}
This pairwise approach coupled with KS-DFT has shown to provide very good accuracy despite its very simple nature that adds basically negligible computational time and, for this reason, it is also employed in the majority of force fields as attractive component of the Lennard-Jones potential.\\ 
Each of the above pairwise terms can be further expanded via a second-order perturbative approach\cite{stone} (in the limit of large inter-atomic separation) to include the higher-order contributions shown in Eq.\eqref{c6c8c10} where $\alpha^j_k (\mathrm{i}\nu)$ is the $k$-pole polarizability at imaginary frequency of atom $j$ (only the single pair $ij$ is considered for the sake of simplicity).\cite{tangdisp}
\begin{equation}
    \label{c6c8c10}
    \begin{split}
    \mathcal{E}_\text{disp}&=-\frac{C_6^{ij}}{R^6_{ij}}-\frac{C_8^{ij}}{R^8_{ij}}-\frac{C_{10}^{ij}}{R^{10}_{ij}}-\dots \\
   C_{2n}^{ij} &=\frac{(2n-2)!}{2\pi}\sum_{h=1}^{n-2}\frac{1}{(2h)!(2k)!}\int_0^\infty d\nu\alpha^i_h(\mathrm{i}\nu) \alpha^j_k(\mathrm{i}\nu) ~~~,~~~k=n-h-1
        \end{split}
\end{equation}
While providing the correct long-range asymptotic limit, $C_6$ terms alone are not enough to describe the short- and medium-range dispersion effects and these higher-order terms have proven to increase significantly interaction energies near equilibrium regions as well as condensed phase properties obtained with molecular dynamics simulations based on force field potentials.\cite{daan}\\
Regardless of the higher-order terms inclusion discussed, this pairwise approach completely neglects the collective many-body dispersion (MBD) effects inherited from the intrinsic quantum mechanical nature of long-range electron correlation and their relevance has been proven in modelling extended systems, supramolecular complexes and proteins in solutions.\cite{mbd_extended,mbd_qmc,plasmonic1,plasmonicmbd}\\ 
These non-additive dispersion effects have been modelled via a set of coupled fluctuating dipoles\cite{mbd_cfd,mbd_cfd2} as well as quantum Drude oscillators\cite{qdo}. In recent years, Tkatchenko, DiStasio Jr. and Ambrosetti have proposed the MBD@rsSCS introducing the range-separation of the self-consistent screening (rsSCS) of atomic  polarizabilities based on Tkatchenko-Scheffler volume rescaling that relies on Hirshfeld's molecular electron density partitioning.\cite{mbd_rsscs} This many-body dispersion approach is particularly elegant as it fully relies on \emph{ab initio}-derived parameters (atomic polarizabilities) except for the unique range separation parameter governing the coupling to the chosen KS-DFT method while providing chemical accuracy.\cite{mbd_rsscs} A fractionally ionic variant of the MBD@rsSCS model (MBD@rsSCS/FI) has also been proposed by Gould \emph{et al.}  where iterative Hirshfeld (HI) density partitioning together with a charge-dependent atomic polarizability approach are embraced,\cite{fracpol} making the partitioning more suitable than the original Hirshfeld scheme in the treatment of ionic compounds.\cite{hi_ionic}\\
Newer developments in the MBD modelling include the nonlocal many-body dispersion method (MBD-NL) of Hermann and Tkatchenko\cite{mbd-nl} where the MBD approach is combined to the Vydrov and Van Voorhis (VV),
polarizability functional\cite{vv} as well as our recently proposed DNN-MBD model where atomic polarizabilities are obtained via deep neuronal network (DNN) model, thus bypassing the explicit electron density partitioning.\cite{dnnmbd} \\
Being based on the interaction among coupled fluctuating dipoles, the MBD model (rsSCS is implicitly assumed and from now dropped) and its variations represent the many-body counterpart of Eq.\eqref{london}, where the effect of higher-order fluctuating multipoles are neglected, altough Massa \emph{et al.} have recently proposed a beyond-dipole MBD model based on the Random Phase Approximation (RPA) formalism.\cite{massabeyond}\\
In the same line, this work independently generalizes the MBD model to coupled fluctuating dipole and quadrupoles in order to improve the description of short and mid-range many-body dispersion. Particularly, we will show how atomic quadrupole polarizabilities can be recursively derived from dipole ones, thus without the need on introducing further parameters. Furthermore, in doing so we will adopt our recently proposed DNN model approach that completely bypasses the explicit electron density partitioning. The outcoming density-free DNN-MBDQ model exhibits improved accuracy especially near equilibrium regions without the inclusion of any additional parameter, compared to other  MBD-based models.\\

\subsection{Theory}
The correlation energy $\mathcal{E}_c$ for a system of interacting electrons can be rigorously expressed with the adiabatic connection fluctuation-dissipation formula in Eq.\eqref{acfd}, $v$ being the Coulomb potential $\lvert \mathbf{r}-\mathbf{r}' \rvert^{-1} $ whose coupling strength is governed by $\lambda$ while $\chi_\lambda$ and $\chi_0$ are the interacting and non-interacting response functions where the latter can be evaluated from a set of single particle orbitals $\phi_i(\mathbf{r})$ with their corresponding energies $\epsilon_i$ and occupation numbers $f_i$ via the Adler-Wiser formalism, Eq.\eqref{aw}.\cite{adler,wiser}

\begin{equation}
\label{acfd}
\begin{split}
\mathcal{E}_c&=-\frac{1}{2\pi}\int_0^\infty d\nu\int_0^1 d\lambda \trace{[\chi_\lambda(\mathbf{r},\mathbf{r}',\mathrm{i}\nu)v(\mathbf{r},\mathbf{r}')-\chi_0(\mathbf{r},\mathbf{r}',\mathrm{i}\nu) v(\mathbf{r},\mathbf{r}')]}\\
\end{split}
\end{equation}
\begin{equation}
\label{aw}
\begin{split}
\chi_0(\mathbf{r},\mathbf{r}',\mathrm{i}\nu) &=\sum_{i,j}(f_i-f_j) \frac{\phi_i^*(\mathbf{r})\phi_i(\mathbf{r'})\phi_j^*(\mathbf{r'})\phi_j(\mathbf{r})}{\epsilon_i-\epsilon_j +\mathrm{i}\nu}
\end{split}
\end{equation}
By defining $\chi_\lambda(\mathbf{r},\mathbf{r}',\mathrm{i}\nu)$ via the self-consistent screening Dyson equation, Eq.\eqref{dyson}, and by setting the exchange-correlation kernel $f^\lambda _\text{xc}=0$ as for the RPA\cite{bohm_rpa}, the correlation energy in Eq.\eqref{acfd} assumes the form in Eq.\eqref{xrpa} where the last equality is obtained by anaytical integration over $\lambda$.\cite{tkatchenkorpa}
\begin{equation}
    \label{dyson}
    \chi_\lambda=\chi_0+\chi_0(\lambda v+f^\lambda _\text{xc})\chi_\lambda
\end{equation}
\begin{equation}
    \label{xrpa}
   \begin{split}
      \mathcal{E}_c^\text{RPA}&=-\frac{1}{2\pi}\int_0^\infty d\nu \int_0^1 \frac{d\lambda}{\lambda} \trace{\biggl[\frac{(\lambda v \chi_0)^2}{1-\lambda v \chi_0}\biggr]} \\
      &=\frac{1}{2\pi}\int_0^\infty d\nu   \trace{[\ln(1-\chi_0v)+\chi_0v]}
   \end{split}
\end{equation}

In virtue of its generality, the framework here discussed for the case of electronic correlation, can be used in different contexts and in particular in connection with the many-body dispersion model where the target energy arises from the correlation of coupled fluctuating dipoles.\cite{tkatchenkorpa} In this case the response function assumes the form of the frequency-dependent atom-in-molecule (AIM) isotropic dipole polarizability $\boldsymbol{\alpha}^\mu_i(\mathrm{i}\nu)$ localized at the atomic position $R_i$ while the Coulomb potential $v(\mathbf{r},\mathbf{r}')$ is replaced by the (properly damped) dipole-dipole interaction tensor $\mathbf{T}^\text{LR}_{ij}=\mathbf{T}^\text{LR,$\mu\mu$}_{ij}$.\\
Eq.\eqref{xrpa} (last equality) is thus written in terms of the MBD model's quantities as shown in Eq.\eqref{rpambd}, $\mathbf{A}(\mathrm{i}\nu)$ being the $(3N,3N)$ diagonal superpolarizability matrix having atomic polarizabilities as elements and $\mathbf{T}^\text{LR}$ being the full dipole-dipole interaction tensor excluding self interactions ($\mathbf{T}_{ii}^\text{LR}=\mathbf{0}$). 
We note that Eq.\eqref{rpambd} does not include the $\chi_0v$ term appearing in Eq.\eqref{xrpa} due to the traceless property of the $\mathbf{A}(\mathrm{i}\nu)\mathbf{T}^\text{LR}$ matrix product.
\begin{equation}
    \label{rpambd}
    \begin{split}
    \mathbf{A}(\mathrm{i}\nu)=&\text{Diag}[\boldsymbol{\alpha}^\mu_1(\mathrm{i}\nu),\boldsymbol{\alpha}^\mu_2(\mathrm{i}\nu),\dots,\boldsymbol{\alpha}^\mu_N(\mathrm{i}\nu)]\\
\mathcal{E}_\text{MBD}=&\frac{1}{2\pi}\int_0^\infty d\nu   \trace{[\ln(\mathbf{I}-\mathbf{A}(\mathrm{i}\nu)\mathbf{T}^\text{LR})]} 
   \end{split}
\end{equation}
For MBD models based on coupled fluctuating dipoles, the RPA formula in Eq.\eqref{rpambd} does not represent the most efficient strategy to compute correlation energies solution as the problem can be equivalently solved via exact diagonalization of the MBD potential matrix for a system of coupled quantum harmonic oscillators.\cite{tkatchenkorpa}
However, the power and generality of the RPA formulation of the MBD model allows for the generalization of the model and the inclusion of higher-order polarizabilities as recently discussed by Massa \emph{et al.}.\cite{massabeyond}\\
The matrix form of Eq.\eqref{rpambd} can rather straightforwardly be generalized to higher-order moments gathered in $\mathbf{A}(\mathrm{i}\nu)$ by consistently augmenting the interaction tensor $\mathbf{T}$ with proper interaction blocks. In particular, for a model based on coupled fluctuating dipoles and quadrupoles (Q), the two matrices of dimension ($12N,12N$), N being the number of atoms in the systems, take the form reported in Eq.\eqref{dipquad} where a general out-of-diagonal (thus non-vanishing) block of size ($12,12$) is shown.

\begin{equation}
\label{dipquad}
    \begin{split}
         \mathbf{A}_{kk}(\mathrm{i}\nu)&=\textbf{Diag}[\alpha^\mu_k(\mathrm{i}\nu),\alpha^\mu_k(\mathrm{i}\nu),\alpha^\mu_k(\mathrm{i}\nu),\alpha^Q_k(\mathrm{i}\nu),\alpha^Q_k(\mathrm{i}\nu),\alpha^Q_k(\mathrm{i}\nu),\dots,\alpha^Q_k(\mathrm{i}\nu)]   \\
         &~~~~~~~\\
             \mathbf{T}^\text{LR}_{ik}&=
\begin{pmatrix}
  \begin{matrix}
   \mathbf{T}^\text{LR,$\mu \mu$}_{ik}
  \end{matrix}
  & \rvline &  \mathbf{T}^\text{LR,$\mu Q$}_{ik} \\
\hline
  \mathbf{T}^\text{LR,$\mu Q$}_{ik} & \rvline &
  \begin{matrix}
   & &  & \\
   & ~~~~\mathbf{T}^\text{LR,$Q Q$}_{ik}&  \\
      & &  & 
  \end{matrix}
\end{pmatrix}~~~,~~i\neq k
    \end{split}
\end{equation}
We will now focus our attention on $\mathbf{A}$ and, in particular, on its dipole and quadrupole dynamic polarizability entries being the model's key parameters.\\
As mentioned earlier, a pleasant feature of the MBD model is the fact that it only relies on \emph{ab initio}-derived parameters. In particular, AIM dipole polarizabilities are typically obtained via the volume rescaling approach shown in Eq.\eqref{tspol}, $\alpha_i^\mu$ and $V_i$ being the AIM static dipole polarizability and volume respectively of the i-th atom while the zero superscript denotes free-atom reference quantities.
\begin{equation}\label{tspol}
  \alpha^\mu_i(0)=\biggl(\frac{V_i}{V_i^0}\biggr)\alpha_i^{\mu,0} (0)
\end{equation}
 This volume/polarizability proportionality was discussed by Brinck \emph{et al.}\cite{politzer} and later used by Johson and Becke\cite{johnson_vol} as well as Tkatchenko and Scheffler\cite{tksche} and further adopted in most of the MBD model's variations. Gould discussed a more complicated volume/polarizability relationship\cite{gould_vol} while Szabo \emph{et al.} recently suggested a four-dimensional scaling of dipole polarizability based on single-particle systems' analysis.\cite{szabo_vol} We will, in the present work, stick to the common volume/polarizability proportionality in Eq.\eqref{tspol} as this will allow us to make ready comparisons with our dipole-only DNN-MBD model.\\
The AIM volume $V_i$ is commonly accessed by Hirshfeld partitioning the explicit electron density obtained via the solution of KS equations. Instead, we will adopt the recently proposed efficient and accurate 5-hidden layers DNN model trained with the minimal basis iterative Stockholder atom\cite{mbisa} (MBISA) volumes of approximately 4.6 millions molecules that provides AIM volumes bypassing the electron density partitioning.\cite{dnnmbd} \\
Compared to Hirshfeld partitioning as well as its Iterative HI variant more suitable for ionic systems\cite{hi_ionic}, the MBISA scheme used to generate the training set does not suffer from asymmetric AIM densities arising from the use of free-atom reference densities that, especially for hydrogen atoms, lead to an overestimation of high radial moments, i.e. AIM volumes.\cite{mbisa} A further disadvantage related to the use of free-atom reference densities affecting the HI partitioning arises from the density interpolation for negatively charged atoms. In fact this procedure is, for some anionic species and for all doubly charged anions typically encountered in inorganic oxide clusters, ill-defined.\\
The next step is to find a suitable expression for $\alpha^Q_i(0)$, possibly without the introduction of empirical parameters to preserve the original MBD spirit. At this point, it is possible to relate two consecutive dispersion coefficients by the scaled Starkschall-Gordon relation\cite{c6c8} shown in Eq.\eqref{recc6c8} for the specific case of homonuclear $C^{ii}_6$ and $C^{ii}_8$ coefficients, where $Z_i$ is the atomic number, $\braket{r^n}$ are expectation values or multipole-type moments derived from atomic densities while $\gamma$ scaling factor will be introduced shortly.\cite{c6c8} We note, in passing, that Eq.\eqref{recc6c8} represents a fundamental recursion relation for Grimme's successful DFT-D3 dispersion correction model.\cite{grimme3}
\begin{equation}
\label{recc6c8}    
\begin{split}
C^{ii}_8&=3C^{ii}_6Q_i\\
Q_i&=\gamma \sqrt{Z_i}\frac{\braket{r^4}_i}{\braket{r^2}_i}
\end{split}
\end{equation}
In order to retrieve $\alpha^Q_i(0)$ in terms of known quantities, we can now assume to model atom $i$ via a quantum Drude particle characterized by the frequency $\omega_i$ as this allows us to express $C^{ii}_6$ and $C^{ii}_8$ in terms of dipole and quadrupole polarizabilities, Eqs.\eqref{c6c8drude1} and \eqref{c6c8drude2}.\cite{jonesmartyna}
\begin{equation}
    \label{c6c8drude1}
     C^{ii}_6=\frac{3}{4}\omega_i[\alpha^\mu_i(0)]^2
\end{equation}
\begin{equation}
    \label{c6c8drude2}
        C^{ii}_8=5\omega_i\alpha^\mu_i(0)\alpha^Q_i(0)
\end{equation}
By inserting Eqs.\eqref{c6c8drude1} and \eqref{c6c8drude2} in Eq.\eqref{recc6c8}, it is finally possible to isolate the static AIM quadrupole polarizability in Eq.\eqref{aimquadpol}.
\begin{equation}
    \label{aimquadpol}
    \alpha^Q_i(0)=\frac{9}{20}Q_i\alpha^\mu_i(0)
\end{equation}
The combination of the scaled Starkschall-Gordon relation with a quantum Drude particle expression for the  $C^{ii}_8$ dispersion coefficient was originally adopted by Carter-Fenk \emph{et al.} in their extended symmetry-adapted perturbation theory with MBD (XSAPT+MBD) approach including pairwise dipole-quadrupole dispersion effects.\cite{mbdsapt} In that occasion, the $\gamma$ scaling factor was successfully modelled according to Eq.\eqref{scalingpar}, where $\gamma^0$ was chosen such that noble-gas quadrupole polarizabilities are reproduced and the same strategy is embraced in this work.
\begin{equation}
\label{scalingpar}
    \gamma = \gamma_0 + \exp{(-\sqrt{Z_i}/2)}
\end{equation}
A further discussion about the combination of the Starkschall-Gordon rule with a quantum Drude particle expression is found in the ``Annex'' section at the end of the manuscript.\\
The above discussed recursive relation represents an efficient strategy to derive quadrupole polarizabilities that can be readily applied to any AIM volume partitioning scheme as the atomic volume is the only density-related quantity which, however, in this work is retrieved from an accurate DNN model.\\

The solution of Eq.\eqref{rpambd} requires frequency-dependent atomic polarizabilities while our discussion was, until now, restricted to the zero-frequency (static) case.\\
Dynamical dipole polarizabilities have been successfully modelled via a [0/2] Padé approximant form and Tang \emph{et al.} suggested the same form to be employed also in modelling higher-order multipole dynamical polarizabilities, as shown in Eq.\eqref{pade}, the superscript $M$ denoting a general multipole level and $\omega_k^{M,0}$ the k-th free-atom multipolar-dependent characteristic excitation frequency.\cite{tangdynpol}
\begin{equation}
    \label{pade}
    \alpha^{M}_k(\mathrm{i}\nu)=\frac{\alpha_k^{M,0}(0)}{1-(\mathrm{i}\nu/\omega_k^{M,0})^2}
\end{equation}
In particular, we employ a common $\omega_k^{M,0}$ parameter for both dipole and quadrupole dynamical polarizabilities and we express it as a function of free atom $C_6$ and static dipole polarizability, Eq.\eqref{omegan}.
\begin{equation}
    \label{omegan}
    \omega_k^{Q,0}=\omega_k^{\mu,0}=\frac{4}{3}\frac{C^{0,kk}_6}{[\alpha_k^{\mu,0}(0)]^2}
\end{equation}
This assumption is, however, well defined as it guarantees to model dynamical quadrupole polarizabilities as lower bound quantities while the introduction of \emph{ab initio}-derived scaling factors could be used to improve over this assumption.\cite{tangdynpol}\\

We note that the dipole polarizabilities used in Eq.\eqref{aimquadpol}, can be either taken directly from Eq.\eqref{tspol} or can be further screened by solving Dyson-like self-consistent screening equations.\cite{scs,mbd_rsscs} In this work the latter is chosen and, as a consequence, these screening effects are thus transferred to quadrupole polarizabilities via Eq.\eqref{aimquadpol}.\\
The explicit expressions of the multipole interaction tensor $\mathbf{T}_{ik}^\text{LR}$, is given in Eq.\eqref{texpl} for the $ik$ pair placed at distance $R_{ik}$ and greek letters are used to denote $x,y,z$ Cartesian components. 
\begin{equation}
    \label{texpl}
    \begin{split}
     \mathbf{T}^\text{LR,$\mu \mu$} _{ik,\tau \sigma}&=f_{\mu \mu}(R_{ik},S_{ik}) \nabla^2 _{ik,\tau \sigma} \biggl( \frac{1}{R_{ik}} \biggr)  \\
     \mathbf{T}^\text{LR,$\mu Q$} _{\tau \sigma \theta}&=f_{\mu Q}(R_{ik},S_{ik}) \nabla^3 _{ik,\tau \sigma \theta}\biggl( \frac{1}{R_{ik}} \biggr)  \\
     \mathbf{T}^\text{LR,$Q Q$} _{ik,\tau \sigma \theta \lambda}&=f_{QQ}(R_{ik},S_{ik}) \nabla^4 _{ik,\tau \sigma \theta \lambda}\biggl( \frac{1}{R_{ik}} \biggr)  \\
    \end{split}
\end{equation}
The general form of the Fermi damping function $f_{MM'}(R_{ik},S_{ik})$ (${MM'}$ denoting $\mu \mu$, $\mu Q$ or $QQ$) is defined in Eq.\eqref{fermi} where $R_i^\text{vdw}$ represents the AIM van der Waals radius here modelled as the ratio between the AIM dipole (screened) polarizability and the free atom one while $\beta$ is the parameter modulating the range-separation of the multipole interaction tensor. Finally the $l_{MM'}$ term, represents a multipole-dependent universal factor that ensures a stronger damping as the multipole order increases ($l_{\mu \mu} < l_{\mu Q} < l_{QQ}$), in line to the behavior of the popular Tang and Toennies damping functions based on the incomplete gamma function.\cite{ttdamp}
\begin{equation}
    \label{fermi}
    \begin{split}
        f_{MM'}(R_{ik},S_{ik})&=\frac{1}{1+\exp{(-6(R_{ik}/S_{ik}-1))}}\\
        S_{ik}&=\beta l_{MM'}(R_i^\text{vdw}+R_k^\text{vdw})\\
         R_i^\text{vdw}&=\biggl(\frac{\alpha^{\mu}_i(0)}{\alpha_i^{\mu,0}(0)}\biggr)^{1/3}R_i^{0,\text{vdw}}
    \end{split}
\end{equation}
These $l_{MM'}$ terms could in principle be set as parameters and optimized together with $\beta$ to reduce a target energy error. However, to avoid overfitting of these terms we will choose them once and for all regardless of the employed functional.\\
In particular, the $l_{\mu \mu}$ term was set to 1 as this follows the choice undertaken within dipole-only models. On the other hand, $l_{\mu Q}$ and $l_{QQ}$ were chosen after a screening of a few test values, with the constraint being $l_{\mu Q}< l_{QQ}$ as this prevents short range spurious effects. We note in passing that there is clearly room for improvements if a fully data-driven process was embraced in the optimization of these terms.\\
We solve the RPA formula in Eq.\eqref{rpambd} by means of Gauss-Legendre quadrature, nevertheless, more sophisticated techniques such as the Clenshaw-Curtis quadrature\cite{fancyquad,fancyquad2} could also be potentially employed.\\
The DNN-MBDQ model is implemented in the Tinker-HP package\cite{tinkerhp} where the extremely efficient linear-scaling stochatic Lanczos-based DNN-MBD model is also implemented.\cite{mbd_stoch} \\
In order to design an accurate and reliable dispersion-corrected KS-DFT model, both the functional and the dispersion correction must fulfill specific criteria. These requirements are essential for avoiding error cancellation effects in favor of more physically grounded achievement of small errors, as recently discussed by Price \emph{et al.}.\cite{requirements} While the dispersion correction should possibly include higher-order terms as well as many-body effects, the DFT functional should be dispersionless since by adding a dispersion correction to an exchange-correlation functional that somewhat partially includes dispersion, will lead to double counting at shorter ranges, with a consequent loss of accuracy. In fact, in order to reduce this artifact, dispersion corrections are often excessively damped with an inevitable deterioration of the mid-range interactions' description. Among the possible dispersionless functionals, the B86bPBE has proven numerically stable and, coupled with the exchange-hole dipole moment dispersion model\cite{johnson_first,johnson}, has performed well for molecular and material applications.\cite{noncovint} On account of both the higher-order terms and the many-body nature of the proposed DNN-MBDQ dispersion correction, thus fulfilling the above mentioned criteria, it is of interest to probe its performances in connection with the B86bPBE dispersionless functional. Furthermore the DNN-MBDQ dispersion model is also coupled to the common PBE and PBE0 semi-local density functionals and this allows us to make comparisons with different MBD-based dispersion corrections including our recently proposed dipole-only DNN-MBD.\\
Pure energies have been calculated in all cases with Jensen's pcseg-3 basis set as this corresponds \emph{de facto} to the complete basis set limit.\cite{pcseg} \\
\section*{Results}
\begin{figure}[!htb]
\centering
   \includegraphics[width=0.6\linewidth]{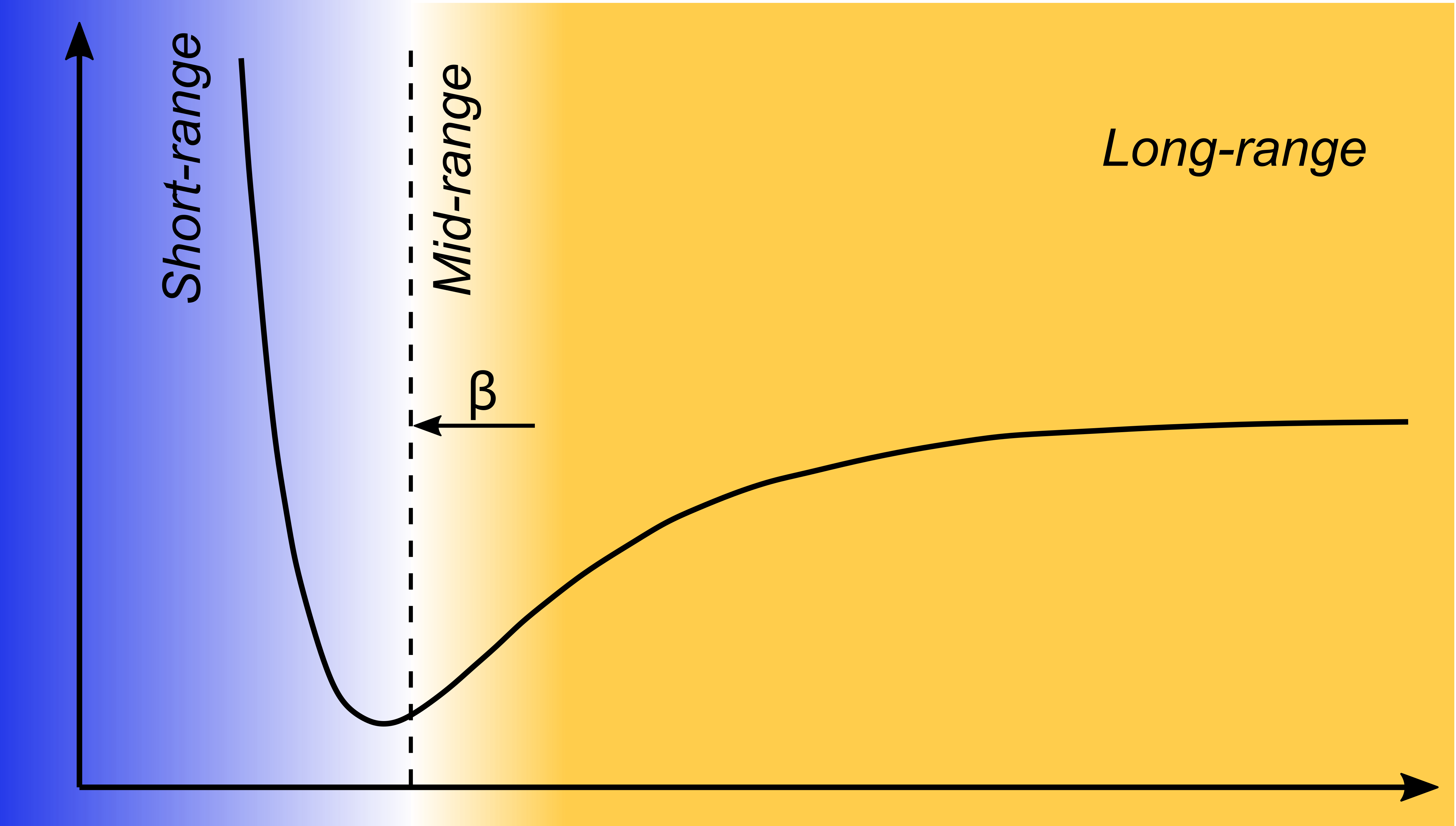}
   \caption{Pictorial representation of the short-, mid- and long-range interaction regimes. The dashed line represents the transition between the short-range part modelled by the KS-DFT functional (blue) and the long-range (yellow) regimes modelled by the DNN-MBD dispersion correction where the introduced quadrupoles improve the modelling of mid-ranges near equilibrium distances. Smaller $\beta$ parameters required by dispersionless functionals have the effect of pushing the yellow zone toward shorter ranges.}
   \label{fig:zones} 
\end{figure}

The S66x8 dataset for noncovalent biologically relevant interactions\cite{s66x8} consisting of 66 dimers placed at 8 different inter-molecular distances (CCSD(T) complete basis set interaction energies) was taken as reference to tune the $\beta$ parameter in Eq.\eqref{fermi} for the three functionals here discussed and the optimal values relative to the DNN-MBDQ model are shown in Table \ref{tab:betas} together with the ones for dipole-only based DNN-MBD. $l_{\mu \mu}$, $l_{\mu Q}$ and $l_{QQ}$ factors are set to 1, 2.8 and 3.0 respectively as these have proven to be good and transferable values for different functionals.
\begin{table}[!htb]
\begin{center}
 \begin{tabular}{||c c c||} 
 \hline
 \textbf{DFT functional} & \textbf{DNN-MBD} & \textbf{DNN-MBDQ} \\ [0.5ex] 
  \hline\hline
    PBE  &   0.75    &    0.82   \\
    PBE0  &   0.77    &    0.83  \\
    B86bPBE & 0.69 & 0.76 \\
\hline
\end{tabular}
\caption{\label{tab:betas}Range-separation $\beta$ parameter values optimized by minimizing the mean absolute relative error for the DNN-MBD and DNN-MBDQ models coupled with the PBE and PBE0 functionals for the S66x8 data set.}
\end{center}
\end{table}
We observe that the PBE0/DNN-MBDQ method requires a larger range-separation parameter compared to the correspondent PBE-based model. This is consistent to what observed also in the PBE/DNN-MBD model due to the PBE0 functional's improved description of short-range exchange-correlation.\\ Moreover, the DNN-MBDQ model requires larger range-separation parameters compared to the DNN-MBD one, and this can be rationalized by virtue of the additional stabilizing dispersion contributions arising from the dipole-quadrupole and quadrupole-quadrupole interactions being the many-body analogues of the pairwise $C_8$ and $C_{10}$ terms in Eq.\eqref{c6c8c10}. We note, in passing, that compared to the original MBD method based on Hirshfeld partitioning for the same functionals (PBE and PBE0), our DNN-MBD model requires smaller $\beta$ values. The reason has to be sought in the relatively smaller AIM volumes predicted by the MBISA method used to build the DNN training set and the consequent effects in AIM  volume-scaled polarizabilities for which less screening is
necessary.
\begin{figure}[!htb]
\centering
   \includegraphics[width=0.5\linewidth]{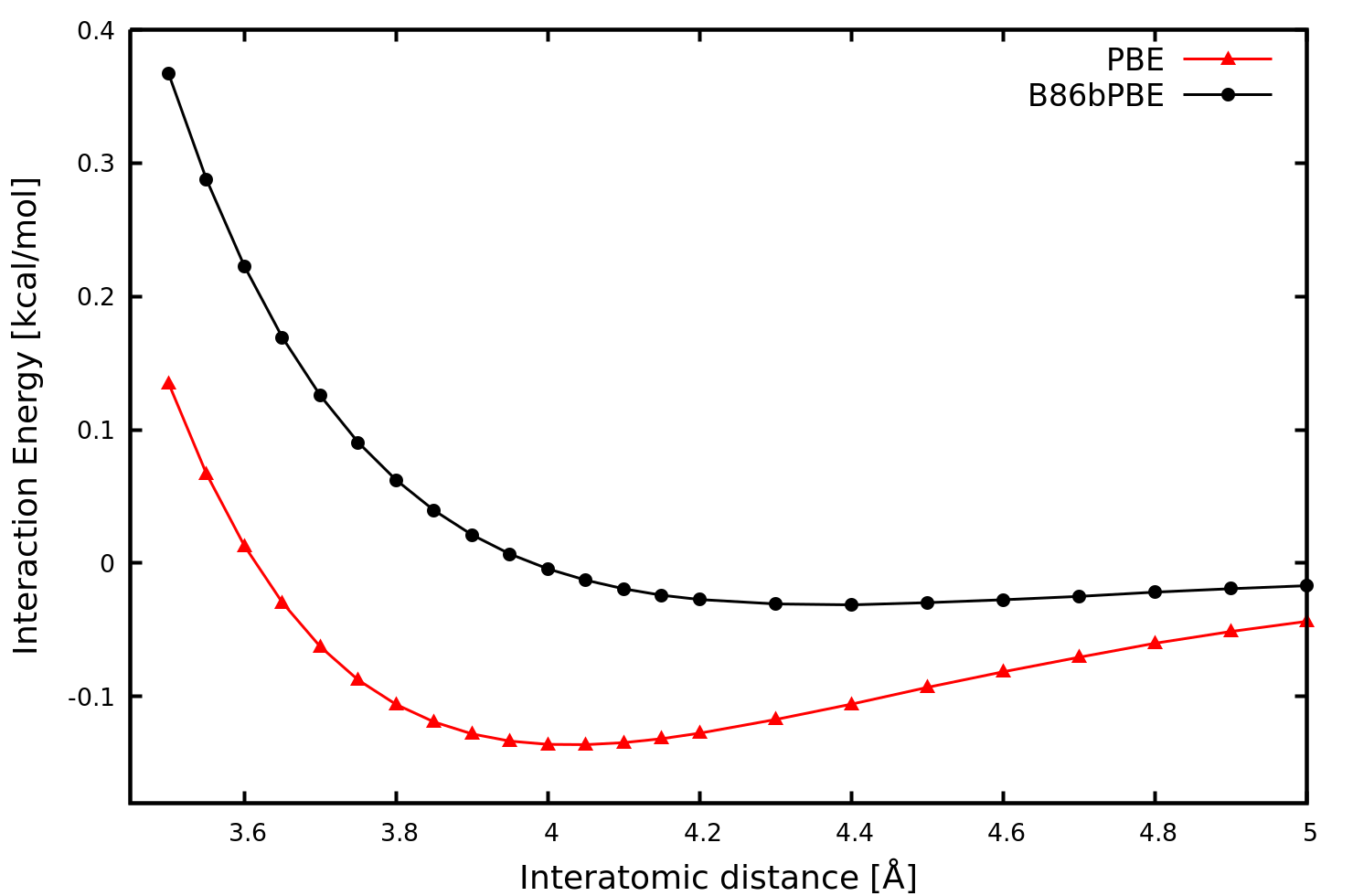}
   \caption{Interaction energy for the argon dimer computed with the PBE and B86bPBE pure functionals (pcseg-3 basis set) as a function of the interatomic distance.}
   \label{fig:ar2} 
\end{figure}
Focusing now on the B86bPBE dispersionless case, we observe that both the DNN-MBD and DNN-MBDQ models require a smaller range-separation parameter (less damping) compared to their PBE/PBE0 counterparts. This is completely in line with the less stabilizing nature of the B86bPBE functional arising from its dispersion-free features as shown in Figure \ref{fig:ar2} where the pure potential energy surface of the argon dimer for the PBE and B86bPBE functionals are shown. In fact, a less damped DNN-MBDQ correction leads to model dispersion at shorter ranges where functionals such as PBE and PBE0 usually involve more damped dispersion corrections to avoid double counting, Figure \ref{fig:zones}.\\
It is now of interest to compare the DNN-MBDQ error for the present dataset with the ones related to different dipole-only MBD models, Table \ref{tab:s66x8}, where the DFT-D3 correction to the PBE functional is also included as a reference method considering its broad use and popularity.
\begin{table}[!htb]
\begin{center}
 \begin{tabular}{||c c c||} 
 \hline
 \textbf{Model} & \textbf{MAE[kcal/mol]} & \textbf{MARE}\% \\ [0.5ex] 
  \hline\hline
    PBE  &   1.55    &    65   \\
    PBE/D3\cite{grimmes66x8}  &   0.44    &    n.a.   \\
    PBE/MBD@rsSCS\cite{mbd_rsscs}  &   0.32    &    10.6   \\
    PBE/MBD@rsSCS/FI\cite{fracpol}  &   0.28    &    9.0   \\    
    PBE/DNN-MBD\cite{dnnmbd} &   0.25    &    9.0   \\   
    \textbf{PBE/DNN-MBDQ}  &   0.24    &    10.5   \\  
    \hline
    PBE0  &   1.48    &    65   \\
    PBE0/MBD@rsSCS\cite{mbd_rsscs} &   0.30    &    9.2   \\
    PBE0/DNN-MBD\cite{dnnmbd} &   0.23    &    6.9   \\
    \textbf{PBE0/DNN-MBDQ} &   0.22    &    8.3   \\
    \hline
    B86bPBE  &   2.01    &   82   \\ 
    B86bPBE/DNN-MBD  &   0.31    &    9.9   \\    
    \textbf{B86bPBE/DNN-MBDQ} &   0.19    &    7.6   \\   


\hline
\end{tabular}
\caption{\label{tab:s66x8}MAE (kcal/mol) and MARE(\%) relative to the S66x8 data set for our density free DNN-MBD and DNN-MBDQ models as well as few other dipole-only MBD models coupled with the PBE and PBE0 functionals. By virtue of its popularity and thus relevance, Grimme's D3 correction is also included. MAE and MARE are computed.}
\end{center}
\end{table}
The proposed DNN-MBDQ model exhibits reduced mean absolute errors, although by a small margin, compared to its dipole-only based DNN-MBD version coupled with both PBE0 and PBE functionals while for the B86bPBE dispersionless functional, the DNN-MBDQ model introduces a more consistent improvement compared to the DNN-MBD and it reaches a notably small value of 0.19 kcal/mol. The MAE of the DNN-MBDQ model coupled to the PBE and PBE0 functionals is always lower than the one relative to other MBD-based models, for which reference data are found in literature, however, except for the B86bPBE functional the DNN-MBDQ model have an slightly higher MARE value. Even in the case of the DNN-MBDQ, dispersion corrected PBE0 provides the smallest error for the considered S66x8 dataset, in line to what observed in the DNN-MBD as well as for other dipole-only MBD models.\\
As anticipated earlier, Massa \emph{et al.} have recently proposed a beyond-dipole MBD model based on the RPA formalism termed MBD+Q@rsSCS, to be coupled to DFT\cite{massabeyond,massabeyond2} and it is thus of interest to compare it with our DNN-MBDQ model. 
In their approach the coupled quantum harmonic oscillators modelling many-body dispersion are parametrized by a mixed approach based on both the Tkatchenko-Scheffler polarizability rescaling and Johnson' exchange dipole model.\cite{johnsonhigher} This differs substantially from our approach where quadrupole polarizabilities are obtained recursively from dipole ones, in turn derived from our DNN-based volume-rescaling scheme. Table \ref{tab:comp} directly compares the MBD+Q@rsSCS and our DNN-MBDQ model based on the complete S66x8 dataset, the S66x7 subset where the shortest inter-molecular distances are disregarded as well as the S66 dataset for both the PBE and PBE0 functionals. 
\begin{table}[!htb]
\begin{center}
 \begin{tabular}{||c c c c||} 
 \hline
 \textbf{Model} &   \textbf{~~S66~~} & \textbf{~~S66x7~~} & \textbf{~~S66x8~~}\\ [0.5ex] 
  \hline\hline
  
    PBE/MBD+Q@rsSCS &   0.45 (9.3\%)    &     0.27 (7.6\%) &     0.34 (12.0\%)   \\
    PBE/DNN-MBDQ   & 0.25 (5.9\%)     &    0.22 (10.3\%)  &    0.24 (10.5\%)    \\    
    \hline
    PBE0/MBD+Q@rsSCS &  0.28 (11.8\%)    &     0.22 (7.1\%) &     0.37 (8.3\%)   \\
    PBE0/DNN-MBDQ   &  0.23  (4.5\%)  &    0.19 (7.8\%) &    0.22 (8.3\%) \\    
\hline
\end{tabular}
\caption{\label{tab:comp}MBD+Q@rsSCS and DNN-MBDQ methods compared in terms of mean absolute errors (kcal/mol) while MARE are reported in brakets for the S66, S66x7 and S66x8 datasets. In the MBD+Q@rsSCS model, two parameters are optimized the number in the damping function against the S66x8 set while for the DNN-MBDQ method a single parameter is optimized.}
\end{center}
\end{table}
In the data sets considered, and for both PBE and PBE0 cases, the DNN-MBDQ exhibits lower MAE and MARE compared to MBD+Q@rsSCS, except for the S66x7 where the MBD+Q@rsSCS model exhibits a lower MARE. In both models, the removal of the shortest inter-molecular distance (S66x7) involves a lowering of the errors compared to the full S66x8 set and this and this denotes the challenge related to modelling mid- to short-range intermolecular distances where dispersion models smoothly turn into local electron correlation governed by the semi-local density functionals. The overall lower errors exhibited by the DNN-MBDQ model can be explained in terms of the differences between two models. In fact, we note that while in the DNN-MBDQ multipole dependent damping functions are employed, the MBD+Q@rsSCS model relies on a single damping function for the different multipole-interaction tensors. At shorter intermolecular distances this may not be optimal as the interaction among higher multipole moments such as quadrupole-quadrupole grows two order of magnitude faster than the dipole-dipole ones and a consistent differentiation of their damping functions is required. While the choice of a multipole-dependent damping functions is essential for multipole interaction tensors obtained from the differentiation of the $1/R$ Coulomb term (Eq.\eqref{fermi}), it becomes less crucial if the intrinsically damped $\erf{(R/\sigma)}/R$ term characterizing the interaction of Gaussian-distributed charge densities of width $\sigma$ is differentiated instead as assumed in the MBD+Q@rsSCS model.\cite{massabeyond} This explains the possibility of using a single Fermi damping function in connection to the MBD+Q@rsSCS model that, although not reaching the accuracy of the DNN-MBDQ model, provides chemically accurate interaction energies. \\
The dipole contribution to dispersion interactions dominates the long-range asymptotic limit while inclusion of quadrupole contributions improves the description at medium ranges near equilibrium distances. The S66x8 is an optimal dataset to optimize the range-separation parameters as it is not biased towards equilibrium distances, however, in virtue or its wide range of inter-molecular distances, it does not represent a fully optimal set to investigate the effects of quadrupole terms. In that regard, results on the S22 dataset, which is composed of model complexes at equilibrium distances, are a better judge of the accuracy of our model since S22 includes more interactions in the full quadrupole (and mixed dipole-quadrupole) operational regime. The S22 validation set thus enables us to better understand the effects of the added quadrupole terms as well as to investigate the transferability of the DNN-MBDQ model by employing the same range-separation parameters reported in Table \eqref{tab:betas}. 
\begin{table}[!htb]
\begin{center}
 \begin{tabular}{||c c c||} 
 \hline
 \textbf{Model} & \textbf{MAE[kcal/mol]} & \textbf{MARE}\% \\ [0.5ex] 
  \hline\hline
    PBE  &   2.66    &    58   \\
    PBE/D3\cite{grimmechemrev}  &   0.48    &    9.9   \\
    PBE/MBD@rsSCS   &   0.49    &    8.9   \\
    PBE/DNN-MBD  &   0.41    &    6.6   \\
    \textbf{PBE/DNN-MBDQ} &   \textbf{0.27}    &    6.0   \\
\hline
    PBE0  &   2.44    &    55   \\
    PBE0/MBD@rsSCS   &   0.55    &    8.5   \\
    PBE0/DNN-MBD   &   0.43    &    5.6   \\
    \textbf{PBE0/DNN-MBDQ}  &   \textbf{0.32}    &    5.9   \\
\hline
    B86bPBE  &   3.29    &   72   \\
    B86bPBE/DNN-MBD   &   0.50    &    10.4   \\
    \textbf{B86bPBE/DNN-MBDQ}  &   \textbf{0.39}    &    9.7   \\
\hline
\end{tabular}
\caption{\label{tab:s22}MAE (kcal/mol) and MARE(\%) relative to the S22 data set\cite{s22sherrill} for some dipole-only based MBD models as well as for our DNN-MBDQ model. For the MBD-based models, the range-separation parameter was optimized against the S66x8 dataset.}
\end{center}
\end{table}
Table \ref{tab:s22} shows the performances for some of the dispersion correction models reported in Table \ref{tab:s66x8} for which results relative to the S22 set are available in literature. Compared to the S66x8, the MAE values reported in Table \ref{tab:s22} for the DNN-MBDQ model coupled to the chosen DFT functionals are higher (although abundantly below the chemical accuracy threshold), however, this is in line with the fact that no $\beta$ optimization was carried out this time. All the PBE, PBE0 and B86bPBE functionals coupled to the DNN-MBDQ model exhibit markedly reduced MAE values compared to the already accurate DNN-MBD dipole-only model while MARE values are not significantly affected. If the DNN-MBD model is compared directly to the MBD@rsSCS model coupled with the PBE and PBE0, the MAE decreases by 45\% and 42\% respectively. These noteworthy performances stem certainly from the additional quadrupolar terms as this is readily observed by a direct comparison between the DNN-MBD and DNN-MBDQ models. The pleasant recursive relation for quadrupole polarizabilities, Eq.\eqref{aimquadpol}, benefits from the DNN model's AIM volumes accuracy arising from the MBISA volumes' data used in the training process whose advantages, compared to the common Hirshfeld partitioning, have been discussed profusely in references and previously recalled in this work.\cite{mbisa,dnnmbd}\\
Among MBD-based models, the DNN-MBDQ coupled with the functionals here considered provides one of the lowest errors for the S22 set without an \emph{ad hoc} parameter optimization. This is rather significant considering that the model fully relies on \emph{ab initio} quantities. 
The accuracy performances of the DNN-MBDQ model could be potentially increased further by adding an extra parameter in the Fermi damping function (Eq.\eqref{fermi}), as well as by performing a complete optimization for the choice of the $l_{MM'}$ terms and future efforts will be spent in this direction.\\ 

\section*{Conclusions}
We propose the DNN-MBDQ model where the density-free/deep learning-aided many-body dispersion model is extended to quadrupolar polarizability terms thanks to a Random Phase Approximation formalism. Quadrupole polarizabilities are recursively retrieved from free-atom dipole polarizabilities and atom-in-molecule volumes modelled via the accurate and transferable neuronal network recently proposed.\cite{dnnmbd} The described stategy can be readily applied to MBD model based on the volume reascaling polarizability partitioning as it does not require any additional electron density-derived quantity. The density-free DNN-MBDQ model implemented in the Tinker-HP package exhibits improved accuracy compared its dipole-only DNN-MBD counterpart as well as reference MBD-based models. In particular, for the widely used S22 data set of dimers placed at equilibrum inter-molecular distances, the DNN-MBDQ shows a remarkable improvement compared to dipole-only reference MBD models.\\
In addition to the common and successful PBE and PBE semi-local functionals, the DNN-MBDQ model is also coupled with the B86bPBE one, to explore the possibility of employing dispersionless functionals, preventing double counting dispersion, especially at shorter ranges. For this case, the errors are abundantly below the chemical accuracy threshold and similar (in some cases even lower) to the ones obtained for the PBE/PBE0 case.\\
We believe that the high accuracy exhibited by the DNN-MBDQ model coupled with semi-local  and dispersionless functionals will be beneficial in pushing Kohn-Sham Density Functional Theory a step closer to post-Hartree Fock reference standards which, due to their high computational cost can not be employed to explore larger systems of relevance in bio- and material modelling. Further work will be therefore dedicated to the evaluation of possible couplings of the DNN-MBDQ approach with existing modern DFT functionals towards generalized chemical accuracy. Such a strategy opens the door to designing accurate, large-scale, energies databases towards various machine learning applications. Finally, the DNN-MBDQ energy correction can be directly added as an \emph{a posteriori} term to pure KS-DFT energies in virtue of the model's density-free features arising from the transferable deep neuronal network employed. 

\begin{acknowledgement}
This work has been funded by the European Research Council (ERC) under the European Union’s Horizon 2020 research and innovation program (grant No 810367), project EMC2 (JPP). Computations have been performed at GENCI (IDRIS, Orsay, France and TGCC, Bruyères le Chatel) on grant no A0070707671.

\end{acknowledgement}

\begin{suppinfo}
SI can be download directly via the Zenodo repository located at the address in reference.\cite{supplementary} SI contains raw PBE, PB0 and B86bPBE interaction energy data for the S66x8 and/or S22 data sets as well as their relative DNN-MBDQ dispersion corrections for the different $\beta$ values discussed. 

\end{suppinfo}
\section*{Annex}
An analytical connection between the Starkschall-Gordon rule and the quantum Drude oscillator (QDO) expression is non-trivial to obtain, however, to verify the consistency of the two expressions, it is possible to evaluate the $C_8/C_6$ proportionality factor for the hydrogen atom, for which $\alpha^Q$, $\alpha^\mu$, $\braket{r^2}$ and $\braket{r^4}$ are analytically available.\cite{Mei_2020,c6c8} \\\\
We start by taking the QDO expression for the $C_8$ coefficient where we introduce $\omega=\frac{4}{3}\frac{C_6}{(\alpha^\mu)^2}$. The resulting expression is the QDO-analogue of the Starkschall-Gordon rule, see equation below where the second equality represents the unscaled Starkschall-Gordon rule for hydrogen.
\begin{equation}
    \nonumber
    \begin{split}
    C_8&=\frac{20}{3}\biggl(\frac{\alpha^Q}{\alpha^\mu}\biggr)C_6=f_{_\text{QDO}}C_6\\
     C_8&=3\biggl(\frac{\braket{r^4}}{\braket{r^2}}\biggr)C_6=f_{_\text{SG}}C_6
    \end{split}
\end{equation}
It is now possible to evaluate $f_{_\text{QDO}}$ and $f_{_\text{SG}}$ from the analytical $\alpha^Q$, $\alpha^\mu$ and $\braket{r^2}$, $\braket{r^4}$ expressions respectively to get $f_{_\text{QDO}}=22.22$ and $f_{_\text{SG}}=22.5$ (atomic units). The $\approx 1\%$ difference among the two factors denotes a marked consistency of the two formulations while the scaling version of the Starkschall-Gordon rule simply improves the results for light elements as discussed by Carter-Fenk, Herbert et al. in connection to their XSAPT+MBD scheme.\cite{mbdsapt}
\bibliography{achemso-demo}

\providecommand{\latin}[1]{#1}
\makeatletter
\providecommand{\doi}
  {\begingroup\let\do\@makeother\dospecials
  \catcode`\{=1 \catcode`\}=2 \doi@aux}
\providecommand{\doi@aux}[1]{\endgroup\texttt{#1}}
\makeatother
\providecommand*\mcitethebibliography{\thebibliography}
\csname @ifundefined\endcsname{endmcitethebibliography}
  {\let\endmcitethebibliography\endthebibliography}{}
\begin{mcitethebibliography}{53}
\providecommand*\natexlab[1]{#1}
\providecommand*\mciteSetBstSublistMode[1]{}
\providecommand*\mciteSetBstMaxWidthForm[2]{}
\providecommand*\mciteBstWouldAddEndPuncttrue
  {\def\EndOfBibitem{\unskip.}}
\providecommand*\mciteBstWouldAddEndPunctfalse
  {\let\EndOfBibitem\relax}
\providecommand*\mciteSetBstMidEndSepPunct[3]{}
\providecommand*\mciteSetBstSublistLabelBeginEnd[3]{}
\providecommand*\EndOfBibitem{}
\mciteSetBstSublistMode{f}
\mciteSetBstMaxWidthForm{subitem}{(\alph{mcitesubitemcount})}
\mciteSetBstSublistLabelBeginEnd
  {\mcitemaxwidthsubitemform\space}
  {\relax}
  {\relax}

\bibitem[Grimme \latin{et~al.}(2016)Grimme, Hansen, Brandenburg, and
  Bannwarth]{grimmechemrev}
Grimme,~S.; Hansen,~A.; Brandenburg,~J.~G.; Bannwarth,~C. Dispersion-corrected
  mean-field electronic structure methods. \emph{Chemical Reviews}
  \textbf{2016}, \emph{116}, 5105--5154\relax
\mciteBstWouldAddEndPuncttrue
\mciteSetBstMidEndSepPunct{\mcitedefaultmidpunct}
{\mcitedefaultendpunct}{\mcitedefaultseppunct}\relax
\EndOfBibitem
\bibitem[Grimme(2004)]{grimme}
Grimme,~S. Accurate description of van der Waals complexes by density
  functional theory including empirical corrections. \emph{Journal of
  Computational Chemistry} \textbf{2004}, \emph{25}, 1463--1473\relax
\mciteBstWouldAddEndPuncttrue
\mciteSetBstMidEndSepPunct{\mcitedefaultmidpunct}
{\mcitedefaultendpunct}{\mcitedefaultseppunct}\relax
\EndOfBibitem
\bibitem[Stone(2013)]{stone}
Stone,~A. \emph{The Theory of Intermolecular Forces}, 2nd ed.; Oxford: Oxford,
  UK, 2013\relax
\mciteBstWouldAddEndPuncttrue
\mciteSetBstMidEndSepPunct{\mcitedefaultmidpunct}
{\mcitedefaultendpunct}{\mcitedefaultseppunct}\relax
\EndOfBibitem
\bibitem[Patil and Tang(1997)Patil, and Tang]{tangdisp}
Patil,~S.~H.; Tang,~K.~T. Multipolar polarizabilities and two- and three-body
  dispersion coefficients for alkali isoelectronic sequences. \emph{The Journal
  of Chemical Physics} \textbf{1997}, \emph{106}, 2298--2305\relax
\mciteBstWouldAddEndPuncttrue
\mciteSetBstMidEndSepPunct{\mcitedefaultmidpunct}
{\mcitedefaultendpunct}{\mcitedefaultseppunct}\relax
\EndOfBibitem
\bibitem[Visscher and Geerke(2020)Visscher, and Geerke]{daan}
Visscher,~K.~M.; Geerke,~D.~P. Deriving a Polarizable Force Field for
  Biomolecular Building Blocks with Minimal Empirical Calibration. \emph{The
  Journal of Physical Chemistry B} \textbf{2020}, \emph{124}, 1628--1636\relax
\mciteBstWouldAddEndPuncttrue
\mciteSetBstMidEndSepPunct{\mcitedefaultmidpunct}
{\mcitedefaultendpunct}{\mcitedefaultseppunct}\relax
\EndOfBibitem
\bibitem[Reilly and Tkatchenko(2013)Reilly, and Tkatchenko]{mbd_extended}
Reilly,~A.~M.; Tkatchenko,~A. Seamless and accurate modeling of organic
  molecular materials. \emph{The Journal of Physical Chemistry Letters}
  \textbf{2013}, \emph{4}, 1028--1033\relax
\mciteBstWouldAddEndPuncttrue
\mciteSetBstMidEndSepPunct{\mcitedefaultmidpunct}
{\mcitedefaultendpunct}{\mcitedefaultseppunct}\relax
\EndOfBibitem
\bibitem[Ambrosetti \latin{et~al.}(2014)Ambrosetti, Alfè, DiStasio, and
  Tkatchenko]{mbd_qmc}
Ambrosetti,~A.; Alfè,~D.; DiStasio,~R.~A.; Tkatchenko,~A. Hard numbers for
  large molecules: toward exact energetics for supramolecular systems.
  \emph{The Journal of Physical Chemistry Letters} \textbf{2014}, \emph{5},
  849--855\relax
\mciteBstWouldAddEndPuncttrue
\mciteSetBstMidEndSepPunct{\mcitedefaultmidpunct}
{\mcitedefaultendpunct}{\mcitedefaultseppunct}\relax
\EndOfBibitem
\bibitem[Ambrosetti \latin{et~al.}(2016)Ambrosetti, Ferri, DiStasio, and
  Tkatchenko]{plasmonic1}
Ambrosetti,~A.; Ferri,~N.; DiStasio,~R.~A.; Tkatchenko,~A. Wavelike charge
  density fluctuations and van der Waals interactions at the nanoscale.
  \emph{Science} \textbf{2016}, \emph{351}, 1171--1176\relax
\mciteBstWouldAddEndPuncttrue
\mciteSetBstMidEndSepPunct{\mcitedefaultmidpunct}
{\mcitedefaultendpunct}{\mcitedefaultseppunct}\relax
\EndOfBibitem
\bibitem[Stöhr and Tkatchenko(2019)Stöhr, and Tkatchenko]{plasmonicmbd}
Stöhr,~M.; Tkatchenko,~A. Quantum mechanics of proteins in explicit water: the
  role of plasmon-like solute-solvent interactions. \emph{Science Advances}
  \textbf{2019}, \emph{5}, eaax0024\relax
\mciteBstWouldAddEndPuncttrue
\mciteSetBstMidEndSepPunct{\mcitedefaultmidpunct}
{\mcitedefaultendpunct}{\mcitedefaultseppunct}\relax
\EndOfBibitem
\bibitem[Langbein(1971)]{mbd_cfd}
Langbein,~D. Microscopic calculation of macroscopic dispersion energy.
  \emph{Journal of Physics and Chemistry of Solids} \textbf{1971}, \emph{32},
  133--138\relax
\mciteBstWouldAddEndPuncttrue
\mciteSetBstMidEndSepPunct{\mcitedefaultmidpunct}
{\mcitedefaultendpunct}{\mcitedefaultseppunct}\relax
\EndOfBibitem
\bibitem[Donchev(2006)]{mbd_cfd2}
Donchev,~A.~G. Many-body effects of dispersion interaction. \emph{The Journal
  of Chemical Physics} \textbf{2006}, \emph{125}, 074713\relax
\mciteBstWouldAddEndPuncttrue
\mciteSetBstMidEndSepPunct{\mcitedefaultmidpunct}
{\mcitedefaultendpunct}{\mcitedefaultseppunct}\relax
\EndOfBibitem
\bibitem[Jones(2010)]{qdo}
Jones,~A. Quantum drude oscillators for accurate many-body intermolecular
  forces. Ph.D.\ thesis, University of Edinburgh, 2010\relax
\mciteBstWouldAddEndPuncttrue
\mciteSetBstMidEndSepPunct{\mcitedefaultmidpunct}
{\mcitedefaultendpunct}{\mcitedefaultseppunct}\relax
\EndOfBibitem
\bibitem[Ambrosetti \latin{et~al.}(2014)Ambrosetti, Reilly, DiStasio, and
  Tkatchenko]{mbd_rsscs}
Ambrosetti,~A.; Reilly,~A.~M.; DiStasio,~R.~A.; Tkatchenko,~A. Long-range
  correlation energy calculated from coupled atomic response functions.
  \emph{The Journal of Chemical Physics} \textbf{2014}, \emph{140},
  18A508\relax
\mciteBstWouldAddEndPuncttrue
\mciteSetBstMidEndSepPunct{\mcitedefaultmidpunct}
{\mcitedefaultendpunct}{\mcitedefaultseppunct}\relax
\EndOfBibitem
\bibitem[Gould \latin{et~al.}(2016)Gould, Lebègue, Ángyán, and
  Bučko]{fracpol}
Gould,~T.; Lebègue,~S.; Ángyán,~J.~G.; Bučko,~T. A fractionally ionic
  approach to polarizability and van der Waals many-body dispersion
  calculations. \emph{Journal of Chemical Theory and Computation}
  \textbf{2016}, \emph{12}, 5920--5930\relax
\mciteBstWouldAddEndPuncttrue
\mciteSetBstMidEndSepPunct{\mcitedefaultmidpunct}
{\mcitedefaultendpunct}{\mcitedefaultseppunct}\relax
\EndOfBibitem
\bibitem[Bučko \latin{et~al.}(2013)Bučko, Lebègue, Hafner, and
  Ángyán]{hi_ionic}
Bučko,~T.; Lebègue,~S.; Hafner,~J.; Ángyán,~J.~G. Improved Density
  Dependent Correction for the Description of London Dispersion Forces.
  \emph{Journal of Chemical Theory and Computation} \textbf{2013}, \emph{9},
  4293--4299, PMID: 26589148\relax
\mciteBstWouldAddEndPuncttrue
\mciteSetBstMidEndSepPunct{\mcitedefaultmidpunct}
{\mcitedefaultendpunct}{\mcitedefaultseppunct}\relax
\EndOfBibitem
\bibitem[Hermann and Tkatchenko(2020)Hermann, and Tkatchenko]{mbd-nl}
Hermann,~J.; Tkatchenko,~A. Density Functional Model for van der Waals
  Interactions: Unifying Many-Body Atomic Approaches with Nonlocal Functionals.
  \emph{Phys. Rev. Lett.} \textbf{2020}, \emph{124}, 146401\relax
\mciteBstWouldAddEndPuncttrue
\mciteSetBstMidEndSepPunct{\mcitedefaultmidpunct}
{\mcitedefaultendpunct}{\mcitedefaultseppunct}\relax
\EndOfBibitem
\bibitem[Vydrov and Van~Voorhis(2010)Vydrov, and Van~Voorhis]{vv}
Vydrov,~O.~A.; Van~Voorhis,~T. Dispersion interactions from a local
  polarizability model. \emph{Phys. Rev. A} \textbf{2010}, \emph{81},
  062708\relax
\mciteBstWouldAddEndPuncttrue
\mciteSetBstMidEndSepPunct{\mcitedefaultmidpunct}
{\mcitedefaultendpunct}{\mcitedefaultseppunct}\relax
\EndOfBibitem
\bibitem[Poier \latin{et~al.}(2022)Poier, Jaffrelot~Inizan, Adjoua, Lagardère,
  and Piquemal]{dnnmbd}
Poier,~P.~P.; Jaffrelot~Inizan,~T.; Adjoua,~O.; Lagardère,~L.; Piquemal,~J.-P.
  Accurate Deep Learning-Aided Density-Free Strategy for Many-Body
  Dispersion-Corrected Density Functional Theory. \emph{The Journal of Physical
  Chemistry Letters} \textbf{2022}, \emph{13}, 4381--4388, PMID: 35544748\relax
\mciteBstWouldAddEndPuncttrue
\mciteSetBstMidEndSepPunct{\mcitedefaultmidpunct}
{\mcitedefaultendpunct}{\mcitedefaultseppunct}\relax
\EndOfBibitem
\bibitem[Massa \latin{et~al.}(2021)Massa, Ambrosetti, and
  Silvestrelli]{massabeyond}
Massa,~D.; Ambrosetti,~A.; Silvestrelli,~P.~L. Many-body van der Waals
  interactions beyond the dipole approximation. \emph{The Journal of Chemical
  Physics} \textbf{2021}, \emph{154}, 224115\relax
\mciteBstWouldAddEndPuncttrue
\mciteSetBstMidEndSepPunct{\mcitedefaultmidpunct}
{\mcitedefaultendpunct}{\mcitedefaultseppunct}\relax
\EndOfBibitem
\bibitem[Adler(1962)]{adler}
Adler,~S.~L. Quantum Theory of the Dielectric Constant in Real Solids.
  \emph{Phys. Rev.} \textbf{1962}, \emph{126}, 413--420\relax
\mciteBstWouldAddEndPuncttrue
\mciteSetBstMidEndSepPunct{\mcitedefaultmidpunct}
{\mcitedefaultendpunct}{\mcitedefaultseppunct}\relax
\EndOfBibitem
\bibitem[Wiser(1963)]{wiser}
Wiser,~N. Dielectric Constant with Local Field Effects Included. \emph{Phys.
  Rev.} \textbf{1963}, \emph{129}, 62--69\relax
\mciteBstWouldAddEndPuncttrue
\mciteSetBstMidEndSepPunct{\mcitedefaultmidpunct}
{\mcitedefaultendpunct}{\mcitedefaultseppunct}\relax
\EndOfBibitem
\bibitem[Bohm and Pines(1953)Bohm, and Pines]{bohm_rpa}
Bohm,~D.; Pines,~D. A Collective Description of Electron Interactions: III.
  Coulomb Interactions in a Degenerate Electron Gas. \emph{Phys. Rev.}
  \textbf{1953}, \emph{92}, 609--625\relax
\mciteBstWouldAddEndPuncttrue
\mciteSetBstMidEndSepPunct{\mcitedefaultmidpunct}
{\mcitedefaultendpunct}{\mcitedefaultseppunct}\relax
\EndOfBibitem
\bibitem[Tkatchenko \latin{et~al.}(2013)Tkatchenko, Ambrosetti, and
  DiStasio]{tkatchenkorpa}
Tkatchenko,~A.; Ambrosetti,~A.; DiStasio,~R.~A. Interatomic methods for the
  dispersion energy derived from the adiabatic connection
  fluctuation-dissipation theorem. \emph{The Journal of Chemical Physics}
  \textbf{2013}, \emph{138}, 074106\relax
\mciteBstWouldAddEndPuncttrue
\mciteSetBstMidEndSepPunct{\mcitedefaultmidpunct}
{\mcitedefaultendpunct}{\mcitedefaultseppunct}\relax
\EndOfBibitem
\bibitem[Brinck \latin{et~al.}(1993)Brinck, Murray, and Politzer]{politzer}
Brinck,~T.; Murray,~J.~S.; Politzer,~P. Polarizability and volume. \emph{The
  Journal of Chemical Physics} \textbf{1993}, \emph{98}, 4305--4306\relax
\mciteBstWouldAddEndPuncttrue
\mciteSetBstMidEndSepPunct{\mcitedefaultmidpunct}
{\mcitedefaultendpunct}{\mcitedefaultseppunct}\relax
\EndOfBibitem
\bibitem[Becke and Johnson(2006)Becke, and Johnson]{johnson_vol}
Becke,~A.~D.; Johnson,~E.~R. Exchange-hole dipole moment and the dispersion
  interaction: High-order dispersion coefficients. \emph{The Journal of
  Chemical Physics} \textbf{2006}, \emph{124}, 014104\relax
\mciteBstWouldAddEndPuncttrue
\mciteSetBstMidEndSepPunct{\mcitedefaultmidpunct}
{\mcitedefaultendpunct}{\mcitedefaultseppunct}\relax
\EndOfBibitem
\bibitem[Tkatchenko and Scheffler(2009)Tkatchenko, and Scheffler]{tksche}
Tkatchenko,~A.; Scheffler,~M. Accurate molecular van der Waals interactions
  from ground-state electron density and free-atom reference data. \emph{Phys.
  Rev. Lett.} \textbf{2009}, \emph{102}, 073005\relax
\mciteBstWouldAddEndPuncttrue
\mciteSetBstMidEndSepPunct{\mcitedefaultmidpunct}
{\mcitedefaultendpunct}{\mcitedefaultseppunct}\relax
\EndOfBibitem
\bibitem[Gould(2016)]{gould_vol}
Gould,~T. How polarizabilities and C6 coefficients actually vary with atomic
  volume. \emph{The Journal of Chemical Physics} \textbf{2016}, \emph{145},
  084308\relax
\mciteBstWouldAddEndPuncttrue
\mciteSetBstMidEndSepPunct{\mcitedefaultmidpunct}
{\mcitedefaultendpunct}{\mcitedefaultseppunct}\relax
\EndOfBibitem
\bibitem[Szab\'o \latin{et~al.}(2022)Szab\'o, G\'oger, Charry, Karimpour,
  Fedorov, and Tkatchenko]{szabo_vol}
Szab\'o,~P.; G\'oger,~S.; Charry,~J.; Karimpour,~M.~R.; Fedorov,~D.~V.;
  Tkatchenko,~A. Four-Dimensional Scaling of Dipole Polarizability in Quantum
  Systems. \emph{Phys. Rev. Lett.} \textbf{2022}, \emph{128}, 070602\relax
\mciteBstWouldAddEndPuncttrue
\mciteSetBstMidEndSepPunct{\mcitedefaultmidpunct}
{\mcitedefaultendpunct}{\mcitedefaultseppunct}\relax
\EndOfBibitem
\bibitem[Verstraelen \latin{et~al.}(2016)Verstraelen, Vandenbrande,
  Heidar-Zadeh, Vanduyfhuys, Van~Speybroeck, Waroquier, and Ayers]{mbisa}
Verstraelen,~T.; Vandenbrande,~S.; Heidar-Zadeh,~F.; Vanduyfhuys,~L.;
  Van~Speybroeck,~V.; Waroquier,~M.; Ayers,~P.~W. Minimal basis iterative
  stockholder: atoms in molecules for force-field development. \emph{Journal of
  Chemical Theory and Computation} \textbf{2016}, \emph{12}, 3894--3912\relax
\mciteBstWouldAddEndPuncttrue
\mciteSetBstMidEndSepPunct{\mcitedefaultmidpunct}
{\mcitedefaultendpunct}{\mcitedefaultseppunct}\relax
\EndOfBibitem
\bibitem[Starkschall and Gordon(1972)Starkschall, and Gordon]{c6c8}
Starkschall,~G.; Gordon,~R.~G. Calculation of Coefficients in the Power Series
  Expansion of the Long‐Range Dispersion Force between Atoms. \emph{The
  Journal of Chemical Physics} \textbf{1972}, \emph{56}, 2801--2806\relax
\mciteBstWouldAddEndPuncttrue
\mciteSetBstMidEndSepPunct{\mcitedefaultmidpunct}
{\mcitedefaultendpunct}{\mcitedefaultseppunct}\relax
\EndOfBibitem
\bibitem[Grimme \latin{et~al.}(2010)Grimme, Antony, Ehrlich, and
  Krieg]{grimme3}
Grimme,~S.; Antony,~J.; Ehrlich,~S.; Krieg,~H. A consistent and accurate ab
  initio parametrization of density functional dispersion correction (DFT-D)
  for the 94 elements H-Pu. \emph{The Journal of Chemical Physics}
  \textbf{2010}, \emph{132}, 154104\relax
\mciteBstWouldAddEndPuncttrue
\mciteSetBstMidEndSepPunct{\mcitedefaultmidpunct}
{\mcitedefaultendpunct}{\mcitedefaultseppunct}\relax
\EndOfBibitem
\bibitem[Jones \latin{et~al.}(2013)Jones, Crain, Sokhan, Whitfield, and
  Martyna]{jonesmartyna}
Jones,~A.~P.; Crain,~J.; Sokhan,~V.~P.; Whitfield,~T.~W.; Martyna,~G.~J.
  Quantum Drude oscillator model of atoms and molecules: Many-body polarization
  and dispersion interactions for atomistic simulation. \emph{Phys. Rev. B}
  \textbf{2013}, \emph{87}, 144103\relax
\mciteBstWouldAddEndPuncttrue
\mciteSetBstMidEndSepPunct{\mcitedefaultmidpunct}
{\mcitedefaultendpunct}{\mcitedefaultseppunct}\relax
\EndOfBibitem
\bibitem[Carter-Fenk \latin{et~al.}(2019)Carter-Fenk, Lao, Liu, and
  Herbert]{mbdsapt}
Carter-Fenk,~K.; Lao,~K.~U.; Liu,~K.-Y.; Herbert,~J.~M. Accurate and Efficient
  ab Initio Calculations for Supramolecular Complexes: Symmetry-Adapted
  Perturbation Theory with Many-Body Dispersion. \emph{The Journal of Physical
  Chemistry Letters} \textbf{2019}, \emph{10}, 2706--2714\relax
\mciteBstWouldAddEndPuncttrue
\mciteSetBstMidEndSepPunct{\mcitedefaultmidpunct}
{\mcitedefaultendpunct}{\mcitedefaultseppunct}\relax
\EndOfBibitem
\bibitem[Tang \latin{et~al.}(1976)Tang, Norbeck, and Certain]{tangdynpol}
Tang,~K.~T.; Norbeck,~J.~M.; Certain,~P.~R. Upper and lower bounds of two‐
  and three‐body dipole, quadrupole, and octupole van der Waals coefficients
  for hydrogen, noble gas, and alkali atom interactions. \emph{The Journal of
  Chemical Physics} \textbf{1976}, \emph{64}, 3063--3074\relax
\mciteBstWouldAddEndPuncttrue
\mciteSetBstMidEndSepPunct{\mcitedefaultmidpunct}
{\mcitedefaultendpunct}{\mcitedefaultseppunct}\relax
\EndOfBibitem
\bibitem[Tkatchenko \latin{et~al.}(2012)Tkatchenko, DiStasio, Car, and
  Scheffler]{scs}
Tkatchenko,~A.; DiStasio,~R.~A.; Car,~R.; Scheffler,~M. Accurate and efficient
  method for many-body van der Waals interactions. \emph{Phys. Rev. Lett.}
  \textbf{2012}, \emph{108}, 236402\relax
\mciteBstWouldAddEndPuncttrue
\mciteSetBstMidEndSepPunct{\mcitedefaultmidpunct}
{\mcitedefaultendpunct}{\mcitedefaultseppunct}\relax
\EndOfBibitem
\bibitem[Tang and Toennies(1984)Tang, and Toennies]{ttdamp}
Tang,~K.~T.; Toennies,~J.~P. An improved simple model for the van der Waals
  potential based on universal damping functions for the dispersion
  coefficients. \emph{The Journal of Chemical Physics} \textbf{1984},
  \emph{80}, 3726--3741\relax
\mciteBstWouldAddEndPuncttrue
\mciteSetBstMidEndSepPunct{\mcitedefaultmidpunct}
{\mcitedefaultendpunct}{\mcitedefaultseppunct}\relax
\EndOfBibitem
\bibitem[Eshuis \latin{et~al.}(2010)Eshuis, Yarkony, and Furche]{fancyquad}
Eshuis,~H.; Yarkony,~J.; Furche,~F. Fast computation of molecular random phase
  approximation correlation energies using resolution of the identity and
  imaginary frequency integration. \emph{The Journal of Chemical Physics}
  \textbf{2010}, \emph{132}, 234114\relax
\mciteBstWouldAddEndPuncttrue
\mciteSetBstMidEndSepPunct{\mcitedefaultmidpunct}
{\mcitedefaultendpunct}{\mcitedefaultseppunct}\relax
\EndOfBibitem
\bibitem[Mussard \latin{et~al.}(2016)Mussard, Rocca, Jansen, and
  Ángyán]{fancyquad2}
Mussard,~B.; Rocca,~D.; Jansen,~G.; Ángyán,~J.~G. Dielectric Matrix
  Formulation of Correlation Energies in the Random Phase Approximation:
  Inclusion of Exchange Effects. \emph{Journal of Chemical Theory and
  Computation} \textbf{2016}, \emph{12}, 2191--2202, PMID: 26986444\relax
\mciteBstWouldAddEndPuncttrue
\mciteSetBstMidEndSepPunct{\mcitedefaultmidpunct}
{\mcitedefaultendpunct}{\mcitedefaultseppunct}\relax
\EndOfBibitem
\bibitem[Lagardère \latin{et~al.}(2018)Lagardère, Jolly, Lipparini, Aviat,
  Stamm, Jing, Harger, Torabifard, Cisneros, Schnieders, Gresh, Maday, Ren,
  Ponder, and Piquemal]{tinkerhp}
Lagardère,~L.; Jolly,~L.; Lipparini,~F.; Aviat,~F.; Stamm,~B.; Jing,~Z.~F.;
  Harger,~M.; Torabifard,~H.; Cisneros,~G.~A.; Schnieders,~M.~J.; Gresh,~N.;
  Maday,~Y.; Ren,~P.~Y.; Ponder,~J.~W.; Piquemal,~J.~P. Tinker-HP: a massively
  parallel molecular dynamics package for multiscale simulations of large
  complex systems with advanced point dipole polarizable force fields.
  \emph{Chem. Sci.} \textbf{2018}, \emph{9}, 956--972\relax
\mciteBstWouldAddEndPuncttrue
\mciteSetBstMidEndSepPunct{\mcitedefaultmidpunct}
{\mcitedefaultendpunct}{\mcitedefaultseppunct}\relax
\EndOfBibitem
\bibitem[Poier \latin{et~al.}(2022)Poier, Lagardère, and Piquemal]{mbd_stoch}
Poier,~P.~P.; Lagardère,~L.; Piquemal,~J.-P. O(N) stochastic evaluation of
  many-body van der Waals energies in large complex systems. \emph{Journal of
  Chemical Theory and Computation} \textbf{2022}, \emph{18}, 1633--1645\relax
\mciteBstWouldAddEndPuncttrue
\mciteSetBstMidEndSepPunct{\mcitedefaultmidpunct}
{\mcitedefaultendpunct}{\mcitedefaultseppunct}\relax
\EndOfBibitem
\bibitem[Price \latin{et~al.}(2021)Price, Bryenton, and Johnson]{requirements}
Price,~A. J.~A.; Bryenton,~K.~R.; Johnson,~E.~R. Requirements for an accurate
  dispersion-corrected density functional. \emph{The Journal of Chemical
  Physics} \textbf{2021}, \emph{154}, 230902\relax
\mciteBstWouldAddEndPuncttrue
\mciteSetBstMidEndSepPunct{\mcitedefaultmidpunct}
{\mcitedefaultendpunct}{\mcitedefaultseppunct}\relax
\EndOfBibitem
\bibitem[Becke and Johnson(2005)Becke, and Johnson]{johnson_first}
Becke,~A.~D.; Johnson,~E.~R. A density-functional model of the dispersion
  interaction. \emph{The Journal of Chemical Physics} \textbf{2005},
  \emph{123}, 154101\relax
\mciteBstWouldAddEndPuncttrue
\mciteSetBstMidEndSepPunct{\mcitedefaultmidpunct}
{\mcitedefaultendpunct}{\mcitedefaultseppunct}\relax
\EndOfBibitem
\bibitem[Becke and Johnson(2007)Becke, and Johnson]{johnson}
Becke,~A.~D.; Johnson,~E.~R. Exchange-hole dipole moment and the dispersion
  interaction revisited. \emph{The Journal of Chemical Physics} \textbf{2007},
  \emph{127}, 154108\relax
\mciteBstWouldAddEndPuncttrue
\mciteSetBstMidEndSepPunct{\mcitedefaultmidpunct}
{\mcitedefaultendpunct}{\mcitedefaultseppunct}\relax
\EndOfBibitem
\bibitem[Johnson(2017)]{noncovint}
Johnson,~E.~R. In \emph{Non-covalent Interactions in Quantum Chemistry and
  Physics}; de~la Roza,~A.~O., DiLabio,~G., Eds.; Elsevier, 2017; pp
  169--194\relax
\mciteBstWouldAddEndPuncttrue
\mciteSetBstMidEndSepPunct{\mcitedefaultmidpunct}
{\mcitedefaultendpunct}{\mcitedefaultseppunct}\relax
\EndOfBibitem
\bibitem[Jensen(2014)]{pcseg}
Jensen,~F. Unifying general and segmented contracted basis sets. Segmented
  polarization consistent basis sets. \emph{Journal of Chemical Theory and
  Computation} \textbf{2014}, \emph{10}, 1074--1085\relax
\mciteBstWouldAddEndPuncttrue
\mciteSetBstMidEndSepPunct{\mcitedefaultmidpunct}
{\mcitedefaultendpunct}{\mcitedefaultseppunct}\relax
\EndOfBibitem
\bibitem[Řezáč \latin{et~al.}(2011)Řezáč, Riley, and Hobza]{s66x8}
Řezáč,~J.; Riley,~K.~E.; Hobza,~P. S66: A Well-balanced database of
  benchmark interaction energies relevant to biomolecular structures.
  \emph{Journal of Chemical Theory and Computation} \textbf{2011}, \emph{7},
  2427--2438\relax
\mciteBstWouldAddEndPuncttrue
\mciteSetBstMidEndSepPunct{\mcitedefaultmidpunct}
{\mcitedefaultendpunct}{\mcitedefaultseppunct}\relax
\EndOfBibitem
\bibitem[Goerigk \latin{et~al.}(2011)Goerigk, Kruse, and Grimme]{grimmes66x8}
Goerigk,~L.; Kruse,~H.; Grimme,~S. Benchmarking Density Functional Methods
  against the S66 and S66x8 Datasets for Non-Covalent Interactions.
  \emph{ChemPhysChem} \textbf{2011}, \emph{12}, 3421--3433\relax
\mciteBstWouldAddEndPuncttrue
\mciteSetBstMidEndSepPunct{\mcitedefaultmidpunct}
{\mcitedefaultendpunct}{\mcitedefaultseppunct}\relax
\EndOfBibitem
\bibitem[Massa \latin{et~al.}(2021)Massa, Ambrosetti, and
  Silvestrelli]{massabeyond2}
Massa,~D.; Ambrosetti,~A.; Silvestrelli,~P.~L. Beyond-dipole van der Waals
  contributions within the many-body dispersion framework. \emph{Electronic
  Structure} \textbf{2021}, \emph{3}, 044002\relax
\mciteBstWouldAddEndPuncttrue
\mciteSetBstMidEndSepPunct{\mcitedefaultmidpunct}
{\mcitedefaultendpunct}{\mcitedefaultseppunct}\relax
\EndOfBibitem
\bibitem[Becke and Johnson(2006)Becke, and Johnson]{johnsonhigher}
Becke,~A.~D.; Johnson,~E.~R. Exchange-hole dipole moment and the dispersion
  interaction: High-order dispersion coefficients. \emph{The Journal of
  Chemical Physics} \textbf{2006}, \emph{124}, 014104\relax
\mciteBstWouldAddEndPuncttrue
\mciteSetBstMidEndSepPunct{\mcitedefaultmidpunct}
{\mcitedefaultendpunct}{\mcitedefaultseppunct}\relax
\EndOfBibitem
\bibitem[Takatani \latin{et~al.}(2010)Takatani, Hohenstein, Malagoli, Marshall,
  and Sherrill]{s22sherrill}
Takatani,~T.; Hohenstein,~E.~G.; Malagoli,~M.; Marshall,~M.~S.; Sherrill,~C.~D.
  Basis set consistent revision of the S22 test set of noncovalent interaction
  energies. \emph{The Journal of Chemical Physics} \textbf{2010}, \emph{132},
  144104\relax
\mciteBstWouldAddEndPuncttrue
\mciteSetBstMidEndSepPunct{\mcitedefaultmidpunct}
{\mcitedefaultendpunct}{\mcitedefaultseppunct}\relax
\EndOfBibitem
\bibitem[Poier \latin{et~al.}(2022)Poier, Lagardère, and
  Piquemal]{supplementary}
Poier,~P.~P.; Lagardère,~L.~L.; Piquemal,~J.-P. {PBE, PBE0, B86bPBE data and
  relative DNN-MBDQ dispersion corrections for S66x8 and S22 data sets.} 2022;
  \url{https://doi.org/10.48550/arXiv.2210.09784}\relax
\mciteBstWouldAddEndPuncttrue
\mciteSetBstMidEndSepPunct{\mcitedefaultmidpunct}
{\mcitedefaultendpunct}{\mcitedefaultseppunct}\relax
\EndOfBibitem
\bibitem[Mei \latin{et~al.}(2020)Mei, Zhou, Zhong, and Qiao]{Mei_2020}
Mei,~X.; Zhou,~W.; Zhong,~Z.; Qiao,~H. Analytical expressions of
  non-relativistic static multipole polarizabilities for hydrogen-like ions*.
  \emph{Chinese Physics B} \textbf{2020}, \emph{29}, 043101\relax
\mciteBstWouldAddEndPuncttrue
\mciteSetBstMidEndSepPunct{\mcitedefaultmidpunct}
{\mcitedefaultendpunct}{\mcitedefaultseppunct}\relax
\EndOfBibitem
\end{mcitethebibliography}

\end{document}


\section{Data set volumes and volumes ratio.}
\begin{figure}[!htb]
\centering
   \includegraphics[width=0.7\linewidth]{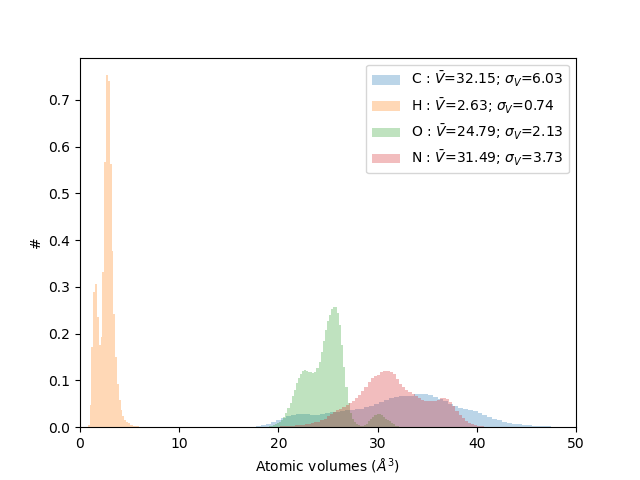}\par
   \includegraphics[width=0.7\linewidth]{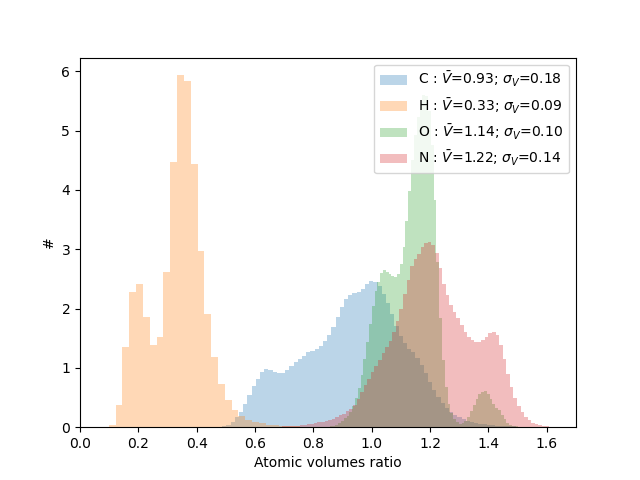}\par
   \caption{Atomic volume (top) and atomic volume ratio (bottom) normalized distributions of the ANI-1 dataset for the C, H, N, O atomic elements.}
   \label{fig:si_anidist} 
\end{figure}

\begin{figure}[!htb]
\centering
   \includegraphics[width=0.7\linewidth]{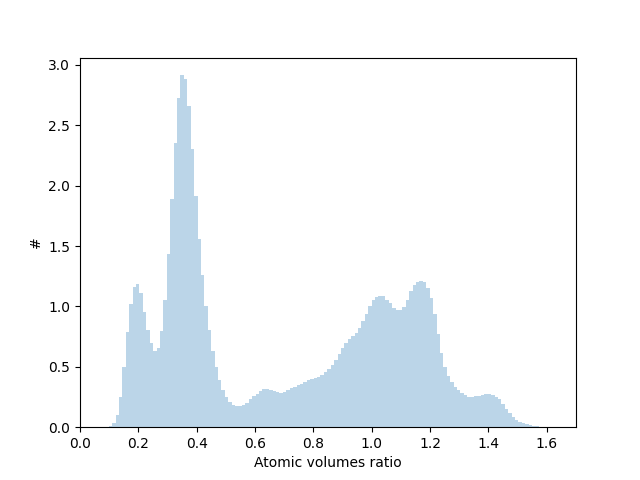}\par
   \includegraphics[width=0.7\linewidth]{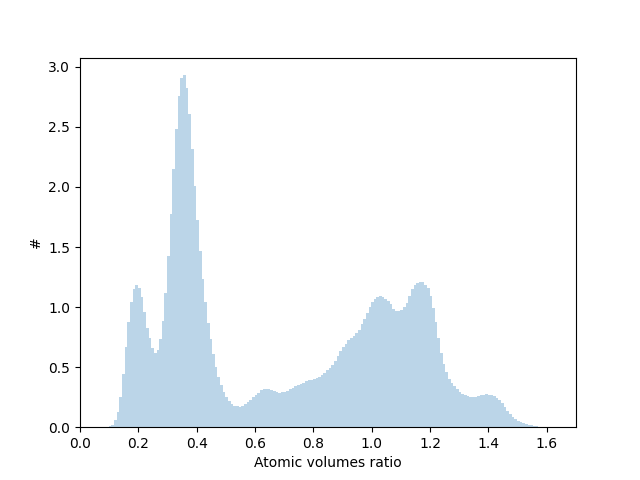}
   \caption{Atomic volume ratio distribution of the training (top) and the validation (bottom) set.}
   \label{fig:si_atomicdist} 
\end{figure}

Distribution of atomic volumes and atomic volumes ratio of the C, H, N and O atomic elements inside the ANI-1 dataset, composed of 4.6 millions of structures are shown in Figure \ref{fig:si_anidist}. Each atomic distribution is normalized to 1. Figure \ref{fig:si_atomicdist} shows the total atomic volumes ratio of the training and validation sets generated from the ANI-1 one. 

\section{Model training and validation errors.}
Below the RMSE evolution as a function of the epoch number is shown for both the training and the validation sets for the four different DNN architectures. In the case of the AIMNET-like model the learning rate reaches the early stopping learning rate within 3000 epoch. The final architectures employed for the final DNN-MBD model is the ANI-AEV 5 layers one.
\begin{figure}[!htb]
\centering
   \includegraphics[width=0.6\linewidth]{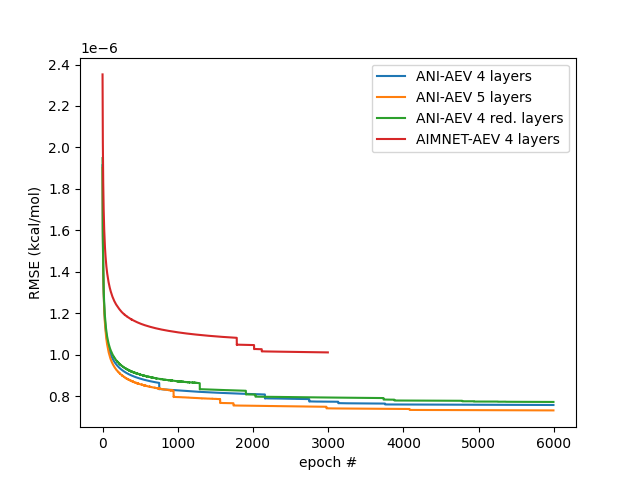}\par
   \includegraphics[width=0.6\linewidth]{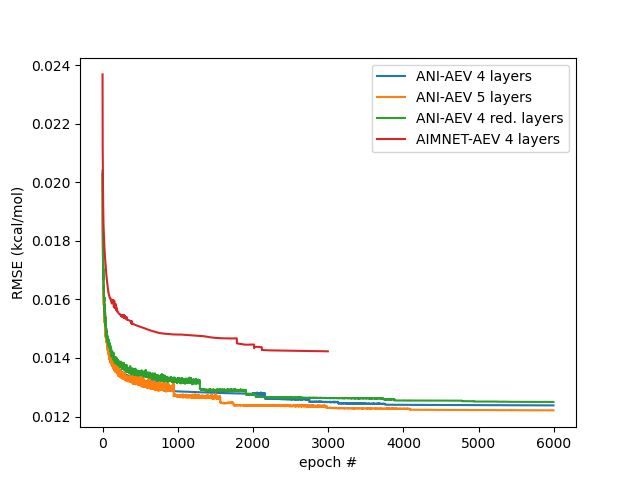}
   \caption{Evolution of the RMSE on volumes ratio w.r.t the number of epoch for the training (top) and the validation (bottom) set.}
   \label{fig:si_learningcurve} 
\end{figure}
The final DNN model and the dataset used for the training can be download directly via the Zenodo repository located at the following address : https://doi.org/10.5281/zenodo.6397639. \\

